\nonstopmode
\documentclass{sig-alternate}%{llncs}

\makeatletter
\def\@ydefthm#1#2[#3]{%
\trivlist
\item[%
\hskip 10\p@
\hskip \labelsep
{\it #2 {\csname the#1\endcsname}.%
\savebox\@tempboxa{{\sc (#3)}}%
\ifdim \wd\@tempboxa>\z@
\ \box\@tempboxa
\fi%
}]%
\ignorespaces
}
\makeatother

\newskip\point\point.8pt
\usepackage{description}

\newif\ifmycolour\mycolourfalse

\newcommand{\It}[1]{\textit{#1}}

\def\renewfont#1#2{\font#1=#2\relax}
\newfont{\itsf}{phvro at 10.25pt}

\usepackage{derivation}
\usepackage{qsymbols}

\usepackage{bbold}
\usepackage{longtable}
\usepackage{mathrsfs}
\usepackage{pstricks}
\usepackage{stmaryrd}
\usepackage{txfonts}
\usepackage{verbatim}
\usepackage{xspace}

\usepackage{mathpazo}
\usepackage{times}
\usepackage{url}

\usepackage{paralist}

\def\OneTo#1{\ensuremath{\overline{#1}}}

\input{commands}
\def\bottom{`W}

\usepackage{fourcolarray}

\newdef{definition}{Definition}
\newdef{Property}{Property}
\newdef{Proposition}{Proposition}
\newdef{Example}{Example}
\newdef{Definition}{Definition}
\newdef{Corollary}{Corollary}
\newenvironment{discussion}{\par {\bf Discussion}:}{\par}

\newtheorem{theorem}{Theorem}%{\begin{Theorem}}{\end{Theorem}}
\newtheorem{lemma}{Lemma}%{\begin{Lemma}}{\end{Lemma}}
\newenvironment{example}{\begin{Example}}{\end{Example}}

\newenvironment{corollary}{\begin{Corollary}}{\end{Corollary}}

\def\TypeRules
{\begin{figure*}[t]
{\small
\[ \begin{array}{rlcrlcrl}
[\rulename{t-ass}]: & 
\Inf [ { \fieldTable(\EC, \Clss[D], \Fld) = \Clss } ]
	{\typeAss{\mvTypeEnvironment}{\Expr}{\Clss[D]}
	 \quad
	 \typeAss{\mvTypeEnvironment}{\Expr' }{\Clss}
	}
	{\typeAss{\mvTypeEnvironment}{\Expr . \Fld = \Expr' }{\Clss[D]} }
&\quad&
[\rulename{t-null}]: &
\Inf[\mbox{\Clss valid in \EC}] {\typeAss{\mvTypeEnvironment}{\Null[\Clss]}{\Clss} }
&\hspace*{2cm}& 
[\rulename{t-var}] : & 
\Inf [\mvTermVar{:}\Clss \ele \mvTypeEnvironment]
	{\typeAss{\mvTypeEnvironment}{\mvTermVar}{\Clss} }
	\\[4mm]
[\rulename{t-fld}]: &
\Inf [ { \fieldTable(\EC, \Clss[D], \Fld) = \Clss } ]
	{ \typeAss{\mvTypeEnvironment}{\Expr}{\Clss[D]} }
	{ \typeAss{\mvTypeEnvironment}{\Expr . \Fld}{\Clss} }
	&&
[\rulename{t-invk}]: &
	\multicolumn{4}{l}{
\Inf [ { \methodTable(\EC, \Clss, \mvMethod) = \Vec[n]{\Clss} \arrow \Clss[D] } ]
	{ \typeAss{\mvTypeEnvironment}{\Expr}{\Clss} 
\quad 
\typeAss{\mvTypeEnvironment}{\Expr_i}{\Clss_i} ~(\forall i \ele \n) }
	{ \typeAss{\mvTypeEnvironment}{\Expr . \mvMethod(\Vec[n]{\Expr})}{\Clss[D]} }
}
	\\[4mm] 
[\rulename{t-sub}]: & 
\Inf [\Clss' \subtypeOf \Clss]
	{ \typeAss{\mvTypeEnvironment}{\Expr}{\Clss' } }
	{ \typeAss{\mvTypeEnvironment}{\Expr}{\Clss}}
	&&
[\rulename{t-new}]: &
	\multicolumn{4}{l}{
\Inf [\fieldListLookup\,(\EC, \Clss) = \mbox{$\Vec[n]{\Fld}$} 
	\And
	\fieldTable(\EC, \Clss, \Fld_i) = \Clss_i~(\forall i \ele \n)]
	{\typeAss{\mvTypeEnvironment}{\Expr_i}{\Clss_i} \quad ~(\forall i \ele \n) }
	{\typeAss{\mvTypeEnvironment}{\keyword{new}\ \Clss(\Vec[n]{\Expr})}{\Clss}} 
}
\end{array} \]
\caption{Type assignment rules} \label{Type Rules}
}
\end{figure*}}

\def\ReductionRules{%
\begin{figure*}
\[ \begin{array}{rclrcl}
		(\NewObject{\Vec[n]{\Expr}}).\Fld_i &\arrow& \Expr_i, & \fieldListLookup(\EC, \Clss) &=& \Vec[n]{\Fld} \And i \ele \overline{n};
	\\[1mm]
(\NewObject{\Vec[n]{\Expr}}).\Fld_{i} = \Expr_{i}' &\arrow& \NewObject{\Expr_1, \ldots, \Expr_i', \ldots, \Expr_n}, &
			\fieldListLookup(\EC, \Clss) &=& \Vec[n]{\Fld} \And i \ele \n;
	\\[1mm]
(\NewObject{\Vec{\Expr}}).\mvMethod(\Vec[n]{\Expr'}) &\arrow& \Expr \substitution{\Vec[n]{\Expr'/\mvTermVar}, \NewObject{\Vec{\Expr}}/\keyword{this}}, &
			\methodLookup(\EC, \Clss, \mvMethod) &=& (\Vec[n]{\mvTermVar}, \Expr).
\end{array} \]
\caption{Reduction rules}\label{reduction rules}
\end{figure*}}

\def\PredicateRules{%
\begin{figure*}[t]
{\small 
\[ \begin{array}{rlcrl}
[\rulename{p-null}]: &
\Inf[\mbox{\Clss valid in \EC}]
	{\PAss{\PEnv}{\Null}{\Clss}{\NullPred}}
&\quad&
[\rulename{p-var}]: &
\Inf[\mvTermVar{:}\Clss\,{:}\,\NPred \ele \PEnv]
	{\PAss{\PEnv}{\mvTermVar}{\Clss}{\NPred}}
\\[3mm]
[\rulename{p-newO}]: &
\Inf{\typeAss{\toTypeEnvironment{\PEnv}}{\NewObject{\Vec{\Expr}}}{\Clss}}
	{\PAss{\PEnv}{\NewObject{\Vec{\Expr}}}{\Clss}{\emptyPredicate}}
&&
[\rulename{p-fld}]: &
\Inf[ { \fieldTable(\EC, \Clss[D], \Fld) = \Clss } ]
	{\PAss{\PEnv}{\Expr}{\Clss[D]}{\mkPredicate{\Fld\,{:}\,\NPred}}}
	{\PAss{\PEnv}{\Expr . \Fld}{\Clss}{\NPred}}
\\[4mm]
[\rulename{p-subT}]: &
\Inf [\Clss' \subtypeOf \Clss \And \NPred \ele \Lang(\Clss) ]
	{\PAss{\PEnv}{\Expr}{\Clss' }{\NPred}}
	{\PAss{\PEnv}{\Expr}{\Clss}{\NPred}}
&&
[\rulename{p-top}]: &
\Inf {\typeAss{\toTypeEnvironment{\PEnv}}{\Expr}{\Clss}}
	{\PAss{\PEnv}{\Expr}{\Clss}{\universalPredicate}}
\\[4mm]
[\rulename{p-join}]: &
\Inf [n > 0]
	{\PAss{\PEnv}{\Expr}{\Clss}{\OPred_{i}} \quad (\forall i\ele\n)}
	{\PAss{\PEnv}{\Expr}{\Clss}{\bigPredicateJoin{\Vec[n]{\OPred}}}}
&&
[\rulename{p-seq}]: &
\Inf[\mbox{$\OPred' \subpredicateOf \OPred \And \OPred \ele \Lang(\Clss)$}]
	{\PAss{\PEnv}{\Expr}{\Clss}{\OPred'}}
	{\PAss{\PEnv}{\Expr}{\Clss}{\OPred}}
\\~
\end{array}
\] \[ 
\begin{array}{rlcrl}
[\rulename{p-ass}_1]: &
\Inf [ { \fieldTable(\EC, \Clss, \Fld) = \Clss[D] } ]
	{\PAss{\PEnv}{\Expr}{\Clss}{\OPred} \quad \PAss{\PEnv}{\Expr'}{\Clss[D]}{\NPred}}
	{\PAss{\PEnv}{\Expr . \Fld = \Expr'}{\Clss}{\mkPredicate{\Fld\,{:}\,\NPred}}}
&\quad\quad& % \\[3mm]
[\rulename{p-ass}_2]: &
\Inf [ { \Fld \notin \Vec[n]{\mvLabel}\And\fieldTable(\EC, \Clss, \Fld) = \Clss[D] } ]
	{\PAss{\PEnv}{\Expr}{\Clss}{\Predicate} \quad \typeAss{\toTypeEnvironment{\PEnv}}{\Expr' }{\Clss[D]}}
	{\PAss{\PEnv}{\Expr . \Fld = \Expr' }{\Clss}{\Predicate}}
\\~
\end{array} 
\] \[ 
\begin{array}{rl}
[\rulename{p-invk}]: &
\Inf [{ \methodTable(\EC, {\Clss[D]}, \mvMethod) = {\Vec[n]{\Clss} }\arrow \Clss }]
	{\PAss{\PEnv}{\Expr}{\Clss[D]}{\mkPredicate{\mvMethod\,{:}\,\Pred[`j] \selfTypeOf \Vec[n]{\Pred} \arrow \NPred}} \quad \PAss{\PEnv}{\Expr}{\Clss[D]}{`j} \quad \PAss{\PEnv}{\Expr_i}{\Clss_i}{\Pred_i} \hspace{5pt} (\forall i \ele \n) }
	{\PAss{\PEnv}{\Expr . \mvMethod(\Vec[n]{\Expr})}{\Clss}{\NPred}}
\\[4mm]
[\rulename{p-newF}]: &
\Inf [ \mbox{$ \begin{array}{c} \fieldListLookup\,(\EC, \Clss) = \Vec[n]{\Fld} \And j \ele \n \And %\\ 
\forall\ i \ele \n \QuantificationConditions{\fieldTable(\EC, \Clss, \Fld_i) = \Clss_i } \end{array}$ } ]
	{\PAss{\PEnv}{\Expr_{j}}{\Clss_{j}}{\NPred} \quad \typeAss{\toTypeEnvironment{\PEnv}}{\Expr_i}{\Clss_i} \quad (\forall i \ele \n \Quant [i \neq j] ) }
	{\PAss{\PEnv}{\NewObject{\Vec[n]{\Expr}}}{\Clss}{\mkPredicate{\Fld_{j}\,{:}\,\NPred}}}
\\[4mm]
[\rulename{p-newM}]: &
\Inf [( \methodTable(\EC, \Clss, \mvMethod) = {\Vec[n]{\Clss}} \arrow {\Clss[D]} \And 
\methodLookup\,(\EC, \Clss, \mvMethod) = ({\Vec[n]{\mvTermVar}}, \Expr_{0}) \And
\PEnv' = \{ {\Vec[n]{ \mvTermVar{:}\Clss\,{:}\,\Pred}}, \keyword{this}{:}\Clss\,{:}\,{\Pred[`j]} \} ) 
	]
	{\typeAss{\toTypeEnvironment{\PEnv}}{\NewObject{\Vec{\Expr}}}{\Clss} \quad 
	 \PAss{\PEnv' }{\Expr_{0}}{\Clss[D]}{\NPred}}
	{\PAss{\PEnv}{\NewObject{\Vec{\Expr}}}{\Clss}{\mkPredicate{\mvMethod\,{:}\,\Pred[`j] \selfTypeOf \Vec[n]{\Pred} \arrow \NPred}}}
\\[-3mm] &
\\[-2mm]
\end{array} \]
\caption{Predicate Assignment Rules} \label{Predicate Rules}
}
\end{figure*}
}

\def\CompPred{%

\begin{figure*}
\[\begin{array}{lrcl}
p = \universalPredicate, \NullPred, \emptyPredicate \Then &
	\Comp{\PEnv}{\Expr}{\Clss}{p} 
	& \Iff & 
	\approximationPredicate{\PEnv}{\Expr}{\Clss}{p} \quad 
\\
	\typeAss{\toTypeEnvironment{\PEnv}}{\Expr}{\Clss} \And \fieldTable(\EC, \Clss, \Fld) = \Clss[D] \Then &
	(\Comp{\PEnv}{\Expr}{\Clss}{\mkUPred{\Fld}{\NPred}}
	& \Iff &
	\Comp{\PEnv}{\Expr . \Fld}{\Clss[D]}{\NPred})
\\
	\typeAss{\toTypeEnvironment{\PEnv}}{\Expr}{\Clss} \And \methodTable(\EC, \Clss, \mvMethod) = \Vec[n]{\Clss} \arrow \Clss[D] \Then &
	(\Comp{\PEnv}{\Expr}{\Clss}{\mkUPred{\mvMethod}{\Pred[`j] \selfTypeOf \Vec[n]{\Pred} \arrow \NPred}} 
	& \Iff &
\\
& \multicolumn{3}{r}{ \quad
	(\Comp{\PEnv}{\Expr}{\Clss}{`j} \And \forall\, i \ele \OneTo{n} \Quant[ \Comp{\PEnv}{\Expr_{i}}{\Clss_{i}}{\Pred_{i}} ]
	\Then 
	\Comp{\PEnv}{\Expr . \mvMethod(\Vec[n]{\Expr})}{\Clss[D]}{\NPred})) 
}
\\
&	\forall\, i \ele \OneTo{n}\Quant [ \Comp{\PEnv}{\Expr}{\Clss}{\OPred_{i}} ] & \Iff & \Comp{\PEnv}{\Expr}{\Clss}{\bigPredicateJoin{\Vec[n]{\OPred}}} ~ (n > 0)
\end{array} \]
\caption{Computability predicate} \label{comp pred}
\end{figure*}}

\begin{document}

\pagestyle{plain}

\conferenceinfo{FTfJP}{'09, July 6 2009, Genova, Italy}
\CopyrightYear{2009} 
\crdata{978-1-60558-540-6/09/07}

\title{Semantic Predicate Types and Approximation for \\ Class-based Object Oriented Programming}
\author{
%\alignauthor
Steffen van Bakel 
\qquad 
Reuben N. S. Rowe \\[1mm]
\affaddr{
	Department of Computing, Imperial College London, 180 Queen's Gate, London SW7 2AZ, U.K. } \\[1mm]
\email{{\tt \small \{svb,rnr07\}@doc.ic.ac.uk} }
}
\date{}

\maketitle

\begin{abstract}
	We apply the principles of the intersection type discipline to the study of 
	class-based object oriented programs and; our work follows from a similar approach (in the context of Abadi and Cardelli's $\varsigma$-object calculus) taken by van Bakel and de'Liguoro. We define an extension of Featherweight Java, \theLanguage, and present a \emph{predicate} system which we show to be sound and expressive. We also show that our system provides a semantic underpinning for the object oriented paradigm by generalising the concept of \emph{approximant} from the Lambda Calculus and demonstrating an approximation result: all expressions to which we can assign a predicate have an approximant that satisfies the same predicate. Crucial to this result is the notion of \emph{predicate language}, which associates a family of predicates with a class.
\end{abstract}

\section{Introduction}
\label{section:Introduction}

It was only after the introduction of object oriented programming that attempts were made to place it on the same theoretical foundations as functional programming. The first were based around extending the Lambda Calculus (\LC) \cite{Barendregt} and representing objects as records \cite{Cardelli84,Mitchell90,CardelliMitchell90,FisherEtAl94}. The seminal work of Abadi and Cardelli \cite{AbadiCardelli96} constitutes perhaps the most comprehensive formal treatment of object orientation, and introduces the \VarsigmaCalculus, which formalises the \emph{object-based} programming paradigm. Similar formal models describing \emph{class-based} languages have been developed as well; notable efforts are Featherweight Java \cite{IgarashiPierceWadler99} and its successor Middleweight Java \cite{BiermanParkinsonPitts03}.

An integral aspect of the theory of programming languages is \emph{type theory} which allows for static analysis via abstract reasoning about programs, so that certain guarantees can be given about their behaviour. 
Type theory easily found acceptance within the world of programming, not only through Milner's claim ``\emph{typed programs cannot go wrong}''\footnote{Here `\emph{wrong}' is a semantic value that represents an error state, created when, for example, trying to apply a number to a number.}, but also because static, compile time type analysis allows for efficient code generation, and the generation of efficient code.
The quest for expressive type systems is still ongoing; for example, types with quantifiers \cite{Girard'72,Reynolds'74} as investigated in the early nineties \cite{Mitchell'88,Pierce'93,Pierce'94,Castagna-Pierce'94}, and the \emph{intersection type discipline} ({\ITD}), as first developed in the early 1980s \cite{CoppoDezani,CDV,BCD,vanBakel95} are two good examples of systems which, while undecidable in principle, have found practical application.
% \Note{More examples? Girard, the cube? What exists for Java?} 

{\ITD} generalises Curry's system by allowing more than one type for free and bound variables, grouped in \emph{intersections} via the type constructor $\inter$.
By introducing this extension a system is obtained that is closed under $ `b$-equality: if $\derI B |- M : `s $ and $M\eqb N$, then $\derI B |- N : `s $, making type assignment undecidable.
Intersection systems satisfy a number of strong properties that are preserved even when considering decidable restrictions.
For example, soundness (subject reduction) will always hold, as does the fact that a term that satisfies certain criteria will terminate (has a normal form), or, with different criteria, produce output (has a head-normal form).
The strength of {\ITD} motivated de'Liguoro \cite{deLiguoro'02} to apply the principles of intersection types to object oriented programming, in particular to the Varsigma Calculus.
Over three papers \cite{Bakel-deLiguro-ICTCS'03,Bakel-deLiguro-ICTCS'05,vanBakel-deLiguoro08}, several systems were explored, for various variants of that calculus.
In this work, we aim to follow up on these efforts and apply the principles of intersection types, and the system of \cite{vanBakel-deLiguoro08} specifically, to a formal model of \emph{class-based} object oriented programming; the model we use is based on \cite{IgarashiPierceWadler99}. 
%We find that we would like to use a slightly richer calculus than \cite{IgarashiPierceWadler99}, but that the collection of features in \cite{BiermanParkinsonPitts03} is, at least for the moment, too complex for our purposes. 
%Therefore, we restrict ourselves, essentially, to \emph{Featherweight Java}. 
Having defined the calculus, we will then prove a subject reduction result. 

The goal of our research is to come to a semantics-based or type-based abstract interpretation for object orientation, for which the present paper contains the first steps. 
To be exact, we show the approximation result: any non-trivial predicate assignment for an expression is also achievable for an approximant of that expression, i.e.~a finite rooted segment of its head-normal form. Thus we link semantics and predicates; the head-normal form is assured to exist by the fact that a non-trivial predicate can be assigned.
This then can be used as a basis for abstract interpretation; an analysis that is immediately within reach is that of \emph{termination}, as we will show in this paper. This is certainly not the only one however; one could think of dead code analysis, type and effect systems, strictness analysis, etc. 

While the abstract interpretation of object-oriented languages has certainly been an active topic of research, the majority of approaches taken thus far appear to have concentrated on control-flow and data-flow analysis techniques rather than type-based abstractions \cite{LogozzoCortesi05}; an exception to this is found in \cite{Grothoff06}. Another observation is that work in this area has been centred around issues of optimisation: \cite{JensenSpoto01} presents a \emph{class analysis} of object-oriented programs which may be used to eliminate virtual function calls, \emph{pointer analysis} \cite{RountevEtAl01} generalises class analysis and also allows for the detection of null pointer dereferencing, and other analyses \cite{Logozzo04a,Logozzo04b} have looked at inferring invariants for classes which can be useful in many optimisations such as the removal of checks for array bounds. 
Termination analysis, missing from this list, is covered by our treatment. Such an analysis has been done on Java bytecode \cite{Albert-et.al.'08}, however our system aims at performing such an analysis directly at the level of the object-oriented language rather than its intermediate form.

The normal, class-based type system for our variant of Java is sound, but not expressive enough to come to in-depth analysis of programs; we therefore introduce the additional concept of predicates, which express the functional behaviour of programs, and allow their execution to be traced. We show that the standard (functional) properties hold and, moreover, put in evidence that we have a strong semantic system: we prove an approximation result and characterise head-normalisation and termination. 
The system, being semi-decidable at best, would need to be limited in expressiveness before it can be used for static analysis. 
This notwithstanding, the main properties shown in this paper would hold also for such a restriction. 

\section{Predicate Featherweight Java}
\label{section:LanguageDefinition}

In this section we define our extension of Featherweight Java (\FeatherweightJava), which we call Predicate Featherweight Java (or \theLanguage). \FeatherweightJava \cite{IgarashiPierceWadler99} is a minimal (functional) calculus based on Java \cite{JLS} which expresses the core features of a class-based object oriented programming language (e.g. inheritance, method invocation and field access, object creation). Its compact nature allows proofs of its properties to be correspondingly succinct. As such, it has proved extremely popular as a starting point for formally studying extensions to Java \cite{IgarashiPierce00,ZhaoEtAl03,ErnstEtAl06,DezaniEtAl06,BettiniEtAl07}. The treatment of \FeatherweightJava and its variants in the literature is very comprehensive, and so here we only define the elements of \theLanguage informally, and discuss its departures from \FeatherweightJava.

\begin{definition}[\theLanguage Syntax]
	The syntax of \theLanguage is defined by the following grammar:
	\[\begin{fiveColArray}
%		\textbf{classes} 
		& \mvClassDef & ::= & \keyword{class}\ \Clss\ \keyword{extends}\ \Clss'\,\{\,\Vec{\mvFieldDef}\ \Vec{\mvMethodDef}\,\} \quad (\Clss \not= \ObjectClass) \\
%		\textbf{methods} 
		& \mvMethodDef & ::= & \Clss[D]\ \mvMethod(\Vec{\Clss\ \mvTermVar})\,\{\,\Expr\,\} \\
%		\textbf{field declarations} 
		& \mvFieldDef & ::= & \Clss\ \Fld \\
%		\textbf{expressions} 
		& \Expr & ::= & \mvTermVar \mid \Null \mid \Expr . \Fld \mid \Expr . \Fld = \Expr' \mid \Expr . \mvMethod(\Vec{\Expr}) \mid \NewObject{\Vec{\Expr}} \\[0.5em]
%		\textbf{execution contexts} 
		& \EC & ::= & \Vec{\mvClassDef} \\
%		\textbf{programs} 
		& \mvProgram & ::= & (\EC, \Expr)
	\end{fiveColArray}\]
\end{definition}

The meta-variables \Clss and \Clss[D] range over class names (which, as in \FeatherweightJava, we also use as types); $\mvMethod$ ranges over method names, $\mvField$ over field identifiers, and $\mvTermVar$ over variable names. The set of class names includes the distinguished name $\ObjectClass$, and the set of variables includes the distinguished name \keyword{this}.

In a similar notation to that of \FeatherweightJava we use $\Vec{\Expr}$ to represent a possibly empty sequence of elements (in this particular case, expressions). When necessary, such a sequence may be subscripted with a meta-variable indicating the number of elements it contains, $\Vec[n]{\Expr}$. Notice that elements of a sequence are permitted to be \emph{composite}, as in $\Vec{\Clss\ \mvTermVar}$. Sequence concatenation is represented by $\Vec{\Expr} \cdot \Vec{\Expr'}$, and \emptySequence denotes the empty sequence. We use \OneTo{n} for the set $\{1,\ldots,n\}$, where n is a natural number.

\begin{definition}
An \emph{execution context} is a sequence of class declarations, and a program is a pair of an execution context and an expression to be evaluated. Classes contain a list of fields and a list of methods, the (class) types of which must be declared. As in \FeatherweightJava, the superclass must always be explicitly declared even if it is \keyword{Object}. Methods may take multiple arguments and method bodies consist of a single expression. 
\end{definition}
We call \EC an execution context, rather than a class table, in order to highlight its purely syntactic nature (as opposed to some form of lookup). 

Notice that \theLanguage does \emph{not} explicitly include constructors, as \FeatherweightJava does. We have chosen to elide this feature since in \FeatherweightJava it is merely `syntactic sugar': constructor methods are never \emph{run}, in the same sense that other methods are invoked, and were included to ensure that all valid \FeatherweightJava programs are also valid Java programs. In \theLanguage, we make object constructors \emph{implicit} by requiring (in the type rule for the \keyword{new} keyword) that the types of the expressions that are to be assigned as field values match the types for the fields as defined by the class of the object being created. We also omit the \keyword{return} keyword in method bodies for the same reason. We feel that this simplifies the calculus without diminishing its relevance in any way.

\ReductionRules
\TypeRules

An important difference between \theLanguage and \FeatherweightJava is that we omit cast expressions in \theLanguage, which were included in \FeatherweightJava in order to support the compilation of Featherweight \GenericJava programs to \FeatherweightJava \cite[\textsection 3]{IgarashiPierceWadler99}. Since that is not an objective of our work, and (more importantly) the presence of downcasts makes the system unsound in the sense that well-typed expressions can reduce to expressions containing meaningless (or `stupid') casts, they are omitted. Upcasts are replaced by subsumption rules in the type system. In \theLanguage we also include syntax to represent the $\Null$ value and a field assignment operation. One of the objectives of our research is to lay a foundation for the treatment of a \emph{stateful} model of object oriented programs, of which these two elements are quintessential components. We therefore feel that it behooves us to incorporate them into the model at the earliest opportunity. Indeed, even at the functional level, we find that their inclusion has some interesting (and non-trivial) consequences: our predicate system can be made expressive enough to catch `null pointer dereferences', and field assignment has important implications for the definition of predicate languages in a complete system.% (see \textsection \ref{section:Completeness}).

\begin{definition}
We use the (syntactic) execution context to define a family of standard lookup functions:
\begin{enumerate}%[(\itshape i\upshape)]
	\item $\fieldListLookup(\EC, \Clss) = \Vec{\mvField}$ returns a sequence of the fields defined (and inherited by) class \Clss;
	\item $\methodLookup(\EC, \Clss, \mvMethod) = (\Vec{\mvTermVar}, \Expr)$ returns the body \Expr of the method $\mvMethod$ in class $\Clss$, along with a sequence containing the names of its formal parameters;
	\item $\fieldTable(\EC, \Clss, \Fld) = \Clss[D]$ returns the type of field $\Fld$ in class $\Clss$;
	\item $\methodTable(\EC, \Clss, \mvMethod) = \Vec{\Clss} \rightarrow \Clss[D]$ returns the signature of method $\mvMethod$ in class $\Clss$.
\end{enumerate}
\end{definition}
We explicitly define the lookup functions such that the class $\ObjectClass$ is empty (i.e.~contains no fields or methods). This is safe since the grammar of \theLanguage precludes the existence of a user-defined class called $\ObjectClass$. An execution context induces a standard subtype relation $\subtypeOf$ defined as the transitive closure of the of class extension.

A number of notions and concepts are defined that strongly depend on the current execution context (like reduction, type assignment, and predicate assignment), and which, therefore, should all be subscripted with its name; but since this context is not changed by execution, we will not do so.
As usual, we impose some well-formedness conditions on execution contexts (e.g. all classes must be uniquely named, and the class hierarhy must be acyclic). 
%\begin{comment}:
\begin{inparaenum}[(\itshape i\upshape)]
	\item all classes are uniquely named;
	\item the class hierarchy is \emph{acyclic};
	\item no class declares a field which it also inherits;
	\item if a class declares a method which it also inherits, then the declared signature must match that of the inherited method;
	\item the variable \variable[this] is not used as a formal parameter to any method;
	\item the types declared for fields and in method signatures must correspond to valid classes, as must all classes that are inherited from.
\end{inparaenum}
Notice that i% \end{comment}
%I
n \emph{well-formed execution contexts} we explicitly forbid the redeclaration of an inherited field, but we allow methods declared higher up in the inheritance hierarchy to be overridden (redefined) in a subclass, subject to the condition that the type signature is identical. Such behaviour is a common (perhaps even integral) aspect of the object oriented paradigm and is also present in \FeatherweightJava.

\subsection{Reduction}

We retain the permissive reduction of \FeatherweightJava (rather than restrict it to a call-by-value semantics as in other extensions, e.g. \cite{BettiniEtAl07}) and extend it to handle field assignment. As in \FeatherweightJava, we use $\Expr \substitution{\Vec[n]{\Expr'/\mvTermVar}}$ to denote the expression obtained by replacing any occurrences of the variables $\mvTermVar_1,\,\ldots\,,\mvTermVar_n$ in $\Expr$ by the expressions $\Expr_1,\,\ldots\,,\Expr_n$ respectively. Formally, a reduction relation \oneStepReductionRelation[\mvSmallExecutionContext] is induced for each execution context; however, as mentioned previously, from now on we will assume a fixed execution context.

\begin{definition}[\theLanguage Reduction]
	The one-step reduction relation is defined as the contextual closure of the rules given in Figure~\ref{reduction rules}.

\end{definition}

\subsection{Type System}

The types of \theLanguage are the same as those of \FeatherweightJava; that is, they are induced by the set of classes defined in the execution context, augmented with $\ObjectClass$. We modify the type system of \FeatherweightJava to handle our extra syntax in an obvious way: we introduce extra rules to allow \Null to be assigned any valid type, and ensure that the r-value in a field assignment expression has the expected type. We also introduce a separate subsumption rule.

\begin{definition}
\begin{enumerate}
\item
If a class \Clss is defined in an execution context \EC, then we say it is \emph{valid in} \EC; $\ObjectClass$ is valid in any execution context. 
\item
A type environment is a set of statements of the form $\mvTermVar{:}\Clss$, which is \emph{well formed} when each statement refers to a uniquely named variable \mvTermVar and a valid type \Clss. 
\item
The typing judgement of \theLanguage is written as $\TAss{\TEnv}{\Expr}{\Clss}$ -- where $`G$ and $\EC$ are well formed -- which reads: \Expr has type \Clss in the type environment \TEnv. The rules of the type assignment system are given in Figure \ref{Type Rules}. 
\item
An execution context is \emph{type consistent} if and only if the execution context is well formed and the body of each method can be assigned its declared return type under the type assumptions given for its parameters in the method signature.
\end{enumerate}
\end{definition}

As for \FeatherweightJava, we can show a soundness result for \theLanguage:
\begin{theorem} For type consistent execution contexts if $\typeAss{\mvTypeEnvironment}{\Expr}{\mvExprType}$ and $\oneStepReduction{\Expr}{\Expr'}$ then $\typeAss{\mvTypeEnvironment}{\Expr'}{\mvExprType}$
\end{theorem}

\section{The Predicate System}
\label{section:PredicateSystem}

We now come to describe the first contribution of our work: the predicate system. Our system aims to provide an analysis which is more expressive than the simple type system of \FeatherweightJava: rather than simply guaranteeing global properties of programs, we wish our predicate types to be semantic in nature, and capture runtime properties. We consider the behaviour of an expression (or rather, the object to which the expression evaluates) in terms of the operations that we may perform on it, i.e.~accessing a field or invoking a method. 
%In this respect we are viewing objects as records of fields and methods, an interpretation that is a legacy of the large body of previous work on object calculi. 
We follow in the tradition of intersection types, originally defined as sequences \cite{Coppo-Dezani'78}, however, by treating our predicates as such: a predicate is a sequence of (potentially incomparable) behaviours, from which any specific one can be selected for an expression as demanded by to the context in which it appears. 
We also incorporate the late typing of self, another important feature found in other intersection type systems for object calculi \cite{Barbanera-deLiguoro04,Bakel-deLiguro-ICTCS'03}. This allows for a greater flexibility in the system, permitting us to update an object prior to invoking a method on it.

We begin by defining our predicate types.

\begin{definition}[Predicates]
	Predicates are defined by the following grammar:
	\[\begin{fourColArray}
		\textit{predicates}: & \Pred & ::= & \universalPredicate \mid \NPred \\
		\textit{normal predicates}: & \NPred & ::= & \NullPred \mid \OPred \\
		\textit{object predicates}: & \OPred & ::= & \mkPredicate{\Vec{\mvLabel{:}\mvMemberPred}} \\
		\textit{member predicates}: & \mvMemberPred & ::= & \NPred \mid \Pred[`j] \selfTypeOf \Vec{\Pred} \rightarrow \NPred
	\end{fourColArray}\]
	where the meta-variable \mvLabel ranges over the set of both field identifiers and method names.
\end{definition}

\PredicateRules

Object predicates thus comprise a sequence of statements describing the behaviour of an object. Each statement associates a certain behaviour (described by the member predicate $\mvMemberPred$) with the result of accessing the field or invoking the method labelled $\mvLabel$. In the case of methods, the predicate additionally indicates the \emph{required} behaviour of the receiver ($`j$) and the arguments ($\Vec{\Pred}$). By combining the predicate \NullPred (denoting a null value) with the object predicates we obtain the set of \emph{normal} predicates, so called because they can be assigned to expressions which evaluate to safe normal forms\footnote{The normal forms are safe in the sense that they do not contain null pointer dereferences.}. The predicate constant \universalPredicate (top) is a standard feature taken from the intersection type discipline, and has the role of covering expressions which do not terminate or, more generally, the result of which bears no relevance to the running of the program in that it does not influence the final outcome.
Notice that intersections are implicitly present in the object predicates, since there is no restriction in place on the labels used: a label can occur more than once. This corresponds to the approach of the strict intersection system \cite{Bakel-TCS'92}.

We now define a subpredicate relation and an operation which combines (object) predicates together. At the heart of intersection type assignment lies the ability to introduce an intersection of types and select a single type from an intersection. In our system the join operation facilitates the former (intersection introduction), and the subpredicate relation allows us to perform intersection elimination.

\begin{definition}[Subpredicate Relation]
	The relation \subpredicateOf is defined as the least pre-order on predicates such that:
	\[ \begin{array}{rcl@{\quad}l}
		\NullPred & \subpredicateOf & \universalPredicate \\
		\emptyPredicate & \subpredicateOf & \universalPredicate 
\end{array} 
\quad
\begin{array}{rcl@{\quad}l}
		\forall i \ele \OneTo{n} \,[\, \Predicate[n] & \subpredicateOf & \mkPredicate{\mvLabel_{i}{:}\mvMemberPred_{i}} \,] & \\
		\forall i \ele \OneTo{n} \,[\, \OPred \subpredicateOf \mkPredicate{\mvLabel_{i}{:}\mvMemberPred_{i}} \,] & \Rightarrow & \OPred \subpredicateOf \Predicate[n] & (n \geq 0)
	\end{array} \]
\end{definition}
Again, this corresponds to the type inclusion relation for strict types.

\begin{definition}[Predicate Join]
	The \emph{join} operation is defined on object predicates as follows:
	\[\PJoin{\mkPredicate{\Vec{\mvLabel{:}\mvMemberPred}}}{\mkPredicate{\Vec{\mvLabel'{:}\mvMemberPred'}}} = \mkPredicate{\Vec{\mvLabel{:}\mvMemberPred} \cdot \Vec{\mvLabel'{:}\mvMemberPred'}}\]
	We generalise the join operation to sequences of object predicates as follows:
\[ \begin{array}{rcl@{\quad}l}
		\bigPredicateJoin{\emptySequence} & = & \emptyPredicate %\\
\end{array} 
\quad
\begin{array}{rcl@{\quad}l}
		\bigPredicateJoin{\OPred \cdot \Vec{\OPred}} & = & \PJoin{\OPred}{(\bigPredicateJoin{\Vec{\OPred}})}
	\end{array} \]
\end{definition}

Since the motivating idea behind predicates is to make a statement on the execution of an expression, we define the notion of a predicate language which allows our system to be truly predictive. For example, by defining this notion, we can show that if we derive the predicate $\mkUPred{\Fld}{\NPred}$ for a typed expression $\Expr{:}\Clss$, then the field $\Fld$ will be visible in the class $\Clss$. Moreover, it will be safe to access the field in $\Expr$, and the result will satisfy the predicate $\NPred$.

\begin{definition}[Predicate Language]
	$\Lang(\Clss)$, the \emph{language} of class \Clss is the smallest set of predicates satisfying the following conditions:
	\begin{enumerate}
\item $\top \ele \Lang(\Clss)$, $\NullPred \ele \Lang(\Clss)$ and $\emptyPredicate \ele \Lang(\Clss)$.
\item $\fieldTable(\EC, \Clss, \Fld) = \Clss[D] \Leftrightarrow (\NPred \ele \Lang(\Clss[D]) \Leftrightarrow \mkPredicate{\Fld{:}\NPred} \ele \Lang(\Clss))$.
\item $\methodTable(\EC, \Clss, \mvMethod) = \Vec[n]{\Clss} \rightarrow \Clss[D] \Leftrightarrow$ 
\\ \hspace*{5mm} 
$(\Pred[`j] \ele \Lang(\Clss) \And \forall \, i \ele \OneTo{n} \,[\, \Pred_i \ele \Lang(\Clss_i) \,] \And \NPred \ele \Lang(\Clss[D]) \Leftrightarrow $
\\ \hspace*{10mm} 
$ \mkPredicate{m : \Pred[`j] \selfTypeOf \Vec[n]{\Pred} \arrow \NPred} \ele \Lang(\Clss))$.
\item $ \forall \, i \in \OneTo{n} \,[\, \OPred_i \in \Lang(\Clss) \,]\Leftrightarrow \bigPredicateJoin{\Vec[n]{\OPred}} \in \Lang(\Clss)$.
	\end{enumerate}
\end{definition}

This notion of language plays a crucial role in the approximation result that is presented in the next section.

\CompPred

\begin{definition}
The rules for our predicate assignment system are given in Figure \ref{Predicate Rules}. 
A predicate environment $\PEnv$, which is a set of statements $\mvTermVar{:}\Clss{:}\Pred$, is well formed if each statement refers to a unique variable $\mvTermVar$, a valid type $\Clss$, and a predicate $\Pred \ele \Lang(\Clss)$. 
The judgement $\PAss{\PEnv}{\Expr}{\Clss}{\Pred}$ -- where again $\PEnv$ and $\EC$ are well formed -- asserts that the expression $\Expr$ of type $\Clss$ can be assigned the predicate $\Pred$ using $\PEnv$.
\end{definition}

Some rules are premised by type assignment judgements, which we write using predicate environments instead of type environments ($\TAss{\toTypeEnvironment{\PEnv}}{\Expr}{\Clss}$). Notice that this is more than a simple notational convenience: formally this is a sound extension since each type environment corresponds to a predicate environment in which the predicate information has been discarded.

We can see the predicate system as a Hoare-style system of pre- and post-conditions. For example, the rule $(\textsc{p-fld})$ expresses that if the expression $\Expr$ satisfies the predicate $\mkUPred{\Fld}{\NPred}$, then accessing the field $\Fld$ will satisfy $\NPred$, giving an annotation like
\[ \begin{array}{l}
\texttt{:: pre:} ~ \Expr ~ \texttt{satisfies} ~ \mkUPred{\Fld}{\NPred} \\
\Expr . \Fld \\
\texttt{:: post: } \NPred
\end{array} \]

\label{late}
As a final comment, we return to the issue of late self typing, mentioned earlier in this section. Notice that a method predicate $\mkUPred{\mvMethod}{`j \selfTypeOf \Vec{\Pred} \rightarrow \NPred}$ is derived only for new object expressions\footnote{This is not strictly true, since we might also derive a method predicate for a variable when it is mentioned in the environment.} using the (\Rule{p-newM}) rule, and that no information about this object (save for its type, which allows us to look up the correct method body) is used to derive the self predicate $`j$. It is only at \emph{the point of invocation} that we check the receiver to ensure it satisfies $`j$. This approach differs from the type systems of \cite{AbadiCardelli96} for the \VarsigmaCalculus, where the self reference in the body of a method may only be given a type reflecting the \emph{current} state of the receiver, even though it may be updated later.

% \subsection{Properties}

We now present the main results of the predicate system. 
% Firstly, we see that the predicate system types exactly the set of expressions that the classic type system does.
\begin{theorem}
\begin{enumerate}
\item
$\exists\ \Pred \QuantificationConditions{\PAss{\PEnv}{\Expr}{\Clss}{\Pred}} \Leftrightarrow \typeAss{\toTypeEnvironment{\PEnv}}{\Expr}{\Clss}$.
\item 
$\PAss{\PEnv}{\Expr}{\Clss}{\Pred} \Then \Pred \ele \Lang(\Clss)$.
\item For type consistent execution contexts if $\PAss{\PEnv}{\Expr}{\mvExprType}{\Pred}$ and $\oneStepReduction{\Expr}{\Expr'}$ then $\PAss{\PEnv}{\Expr'}{\mvExprType}{\Pred}$.
\end{enumerate}
\end{theorem}
%The following theorem forms the basis of the predictive nature of the predicate system that we have mentioned previously.
%Most importantly, the predicate system exhibits a \emph{soundness} result, that is, assignable predicates are preserved by the reduction relation.

\section{Approximants for \theLanguage}
\label{section:Approximation}

In this section, we derive an approximation result which can be used as a basis for semantics-based abstract interpretation, or, more directly, a termination analysis of \theLanguage. It also opens the way forward for giving a denotational semantics to our calculus. 

The notion of approximant was first introduced for the $\lambda$-calculus by C. Wadsworth \cite{Wadsworth}. Intuitively, an approximant can be seen as a `snapshot' of a computation, constructed by covering places where computation (reduction) may still take place with the element $\bottom$\footnote{$`W$ is the symbol originally used in \cite{Wadsworth}; more common now is to, as \cite{Barendregt}, use the symbol $\bot$;
since this could be confused with our predicate $\universalPredicate$, we have opted for the old notation.}.
\begin{definition}
We define \emph{approximate} \theLanguage expressions by the following grammar:
\[ \begin{array}{rcl}
	\mvApprExpr & \Coloneqq & \mvTermVar \mid \bottom \mid \Null \mid \mvApprExpr . \Fld \mid \mvApprExpr . \Fld =\mvApprExpr' \mid \mvApprExpr . \mvMethod(\Vec{\mvApprExpr}) \mid \NewObject{\Vec{\mvApprExpr}} 
\end{array} \]
\end{definition}
By extending the notion of reduction so that any field access, field assignment or method invocation on $\bottom$ itself reduces to the expression $\bottom$, we can also define the notion of approximate \emph{normal forms}.

\begin{definition}
Approximate normal forms are defined by the following grammar:
\[ \begin{array}{rcl}
\mvApprNF & \Coloneqq & \mvTermVar \mid \bottom \mid \Null \mid \NewObject{\Vec{\mvApprNF}} \mid 
\\
	&& 
\mvApprNF . \Fld \mid \mvApprNF . \Fld = \mvApprNF' \mid \mvApprNF . \mvMethod(\Vec{\mvApprNF}) \hspace{10pt} (\mvApprNF \neq \bottom, \NewObject{\Vec{\mvApprNF}})
\end{array} \]
\end{definition}
We extend the type and predicate assignment relations to operate over approximate expressions. We add a type assignment rule permitting $\bottom$ to have any valid type, however we do not modify the predicate assignment rules. In particular, this means that $\bottom$ \emph{must} be assigned the predicate $\universalPredicate$.

To formalise the notion of \emph{snapshot}, we define an ordering on approximate expressions:
\begin{definition}
The \emph{direct approximation} relation \directlyApproximates over approximate expressions is defined as the smallest pre-order satisfying:
	\[\begin{array}{rcl}
		\bottom & \directlyApproximates & \Expr \\
		\Expr \directlyApproximates \Expr' & \Then & \Expr . \Fld \directlyApproximates \Expr' . \Fld \\
		\Expr_{1} \directlyApproximates \Expr_{1}' \And \Expr_{2} \directlyApproximates \Expr_{2}' & \Then & \Expr_{1} . \Fld = \Expr_{2} \directlyApproximates \Expr_{1}' . \Fld = \Expr_{2}' \\
		\Expr \directlyApproximates \Expr' \And \Expr_{i} \directlyApproximates \Expr_{i}' \textrm{ for all } i \ele \OneTo{n} & \Then & \Expr . \mvMethod(\Vec[n]{\Expr}) \directlyApproximates \Expr' . \mvMethod(\Vec[n]{\Expr'}) \\
		\Expr_{i} \directlyApproximates \Expr_{i}' \textrm{ for all } i \ele \OneTo{n} & \Then & \NewObject{\Vec[n]{\Expr}} \directlyApproximates \NewObject{\Vec[n]{\Expr'}}
	\end{array}\]
An \emph{approximant} of an expression $\Expr$ is an approximate normal form $\ApprNF$ which directly approximates some expression $\Expr'$ to which $\Expr$ reduces, except for occurrence of $\bottom$ in $\ApprNF$ (so $\ApprNF \directlyApproximates \Expr'$). We write $\approximantsOf{\Expr}$ to denote the set of all the approximants of $\Expr$.
\end{definition}

The following result gives an approximation semantics to \theLanguage, in which we interpret an expression by its set of approximants,$\AECSem[\Expr] = \approximantsOf{\Expr}$.
\begin{lemma}
$\reducesTo{\Expr}{\Expr'} \Then \approximantsOf{\Expr} = \approximantsOf{\Expr'} $
\end{lemma}

As a shorthand notation, we define an \emph{approximation} predicate:
\begin{definition}
$\approximationPredicate{\PEnv}{\Expr}{\Clss}{\Pred} \Iff $

\hfill $\typeAss{\toTypeEnvironment{\PEnv}}{\Expr}{\Clss} \And \exists\, \mvApprNF \ele \approximantsOf{\Expr} \Quant[ \PAss{\PEnv}{\mvApprNF}{\Clss}{\Pred} ]$.
\end{definition}

The approximation result is the following: any expression to which a predicate can be assigned has an approximant with that same predicate. We follow Tait's proof method \cite{Tait} involving a \emph{computability} predicate. 
Space restrictions do not allow us to present the proofs in detail.

\begin{definition}[Computability Predicate]
\label{def:Computability}
The computability predicate is defined inductively over predicates as in Figure~\ref{comp pred}.
\end{definition}
A key step in the proof is to show that computability implies approximation:
\begin{lemma} \label{thm:CompImpliesAppr} \label{lem:ComputableVariables}
\begin{enumerate}
\item
$\Comp{\PEnv}{\Expr}{\Clss}{\Pred} \Then \Appr{\PEnv}{\Expr}{\Clss}{\Pred}$.
\item
$\PAss{\PEnv}{\mvTermVar}{\Clss}{\Pred} \Then \Comp{\PEnv}{\mvTermVar}{\Clss}{\Pred}$.
\end{enumerate}
\end{lemma}
The next step is to formulate a \emph{replacement} lemma, which states that if we replace all the variables in a predicable expression with expressions computable of appropriate predicates, then we obtain a computable expression.

\begin{lemma}[Replacement Lemma]
If $\PAss{\PEnv}{\Expr}{\Clss}{\Pred}$, and \\ there exists $\PEnv'$, $\Vec[n]{\Expr'}$ such that for all $\mvTermVar_i{:}\Clss_i{:}\Pred_i \ele \PEnv$ we have that $\Comp{\PEnv'}{\Expr_i}{\Clss_i}{\Pred_i}$, then $\Comp{\PEnv'}{\Expr \substitution{\Vec[n]{\mvTermVar/\Expr}}}{\Clss}{\Pred}$.
\end{lemma}
Given that all variables are computable of predicates which are assignable to them (Lemma \ref{lem:ComputableVariables}), we can simply replace all the variables in an expression by themselves, and so a corollary of the replacement lemma is that if an expression can be assigned a predicate then it is also computable of that predicate.
\begin{corollary}
$\PAss{\PEnv}{\Expr}{\Clss}{\Pred} \Then \Comp{\PEnv}{\Expr}{\Clss}{\Pred}$.
\end{corollary}
Combining this with Lemma %Theorem 
\ref{thm:CompImpliesAppr} allows us to derive our approximation result:
\begin{theorem}
If $\PAss{\PEnv}{\Expr}{\Clss}{\Pred}$ then there exists $\mvApprNF \ele \approximantsOf{\Expr}$ such that $\PAss{\PEnv}{\mvApprNF}{\Clss}{\Pred}$.
\end{theorem}

While the approximation result shown above is significant in its own right, perhaps of more interest is that it facilitates a \emph{termination analysis} of \theLanguage. We can show that all expressions to which we can assign a \emph{normal} predicate (i.e.~not \universalPredicate) have a \emph{head-normal form}, that is they will reduce to either the null value or an object\footnote{This holds for expressions typed in an empty environment (closed expressions). In general, a head normal form may also comprise a sequence of field accesses, assignment and method invocations on variables.}.
\begin{definition}[Head normal forms]
Head normal forms for \theLanguage are defined by the following grammar:
\[\begin{array}{rcll}
	\HNFExpr & \Coloneqq & \mvTermVar \mid \Null \mid \NewObject{\Vec{\Expr}} \mid 
\\ &&
	\HNFExpr . \Fld \mid \HNFExpr . \Fld =\Expr \mid \HNFExpr . \mvMethod(\Vec{\Expr}) 
	& (\HNFExpr \not= \Null, \NewObject{\Vec{\Expr}})
\end{array}\]
\end{definition}

\begin{theorem}[Termination]
If $\PAss{\PEnv}{\Expr}{\Clss}{\NPred}$ then there exists $\HNFExpr$ such that $\reducesTo{\Expr}{\HNFExpr}$.
\end{theorem}

To illustrate this result, consider the following program:
\begin{example} Take the environment
 \[ \begin{array}{l}
      \keyword{class C extends Object \{} \\
      \hspace{10px} \keyword{C f} \\
      \hspace{10px} \keyword{C m() \{ this.f \}} \\
      \keyword{\}}
 \end{array} \]
\end{example}
Notice that the expression \keyword{(new C(null)).m()} has the approximant \Null (which is also its normal form). We can easily derive \PAss{\emptyset}{\Null}{\className[C]}{\NullPred} using the (\Rule{p-null}) rule. The following derivation shows that we can also assign this predicate to the original expression:
\[ \kern-5mm
 \Inf	{\Inf {\Inf {\Inf {\PAss{\{\variable[this]{:}\className[C]{:}\mkPredicate{\fieldId[f]{:}\NullPred}\}}{\variable[this]}{\className[C]}{\mkPredicate{\fieldId[f]{:}\NullPred}}}
			}
			{\PAss{\{\variable[this]{:}\className[C]{:}\mkPredicate{\fieldId[f]{:}\NullPred}\}}{\variable[this].\fieldId[f]}{\className[C]}{\NullPred}}
		}
		{\PAss{\emptyset}{\keyword{new C(null)}}{\className[C]}{\mkPredicate{\methodName[m] {:} \mkPredicate{\fieldId[f] {:} \NullPred} \selfTypeOf \emptySequence \rightarrow \NullPred}}}
\raise3\RuleH\hbox to 1cm{\kern-25mm
	 \Inf	{\Inf	{\PAss{\emptyset}{\Null}{\className[C]}{\NullPred}}
		}
		{\PAss{\emptyset}{\keyword{new C(null)}}{\className[C]}{\mkPredicate{\fieldId[f] {:} \NullPred}}}
}
\multiput(-10,0)(0,5){8}{.}
	}
	{\PAss{\emptyset}{\keyword{(new C(null)).m()}}{\className[C]}{\NullPred}}
\]

% \IncompletenessEg

\section{What About Completeness?}
\label{sec:Completeness}

While the system we have presented in this paper is sound (exhibits subject reduction), it is not completely expressive since predicable approximants may exist for an expression to which we cannot assign those same predicates. This is a consequence of the fact that our system does not have a subject expansion property (as other intersection type assignment systems do). While this may easily be achieved by discarding our notion of predicate language, doing so would destroy the semantic underpinning of our system (i.e.~the approximation result). The challenge, therefore, is to construct a system with both properties. While we do not offer a comprehensive solution here, we will discuss the underlying reasons for the failure of subject expansion in the presence of predicate languages as we have defined them, and discuss, at an abstract level, the steps that will be required.

This issue goes right to the heart of the object oriented paradigm since the failure of subject expansion in our system lies in the \emph{dynamic dispatch} feature of OO. 

\begin{example} %given in Figure \ref{Fig:Incompleteness}, 
Take the program 
\[ \begin{array}{l}
\begin{array}[t]{l}
\keyword{class Sub extends Object \{} \\
\hspace{10px} \keyword{A upcast(A x) \{ x \}} \\
\keyword{\}} \\
\end{array} 
\\
\begin{array}[t]{l}
\keyword{class A extends Object \{} \\
\hspace{10px} \keyword{A m() \{ this \}} \\
\keyword{\}} \\
\end{array} 
\\
\begin{array}[t]{l}
\keyword{class B extends A \{} \\
\hspace{10px} \keyword{A f} \\
\hspace{10px} \keyword{A m() \{ this.f \}} \\
\keyword{\}} 
\end{array} 
\end{array}
\]
and the run
\[
\begin{array}{lll}
& \keyword{(new Sub()).upcast(new B(new A())).m()} & (1) \\
\oneStepReducesTo & \keyword{(new B(new A())).m()} & (2) \\
\oneStepReducesTo & \keyword{(new B(new A())).f} & (3) \\
\oneStepReducesTo & \keyword{ new A()} & (4)
\end{array}
\] 
\end{example}

We begin by invoking the method \methodName[m] on the receiving expression \keyword{(new Sub()).upcast(new B(new A()))}. By looking at the execution context, we see that the \keyword{upcast} method returns a result of type \className[A]. However, at runtime, the result is actually an object of type \className[B], namely \keyword{new B(new A())}. Thus, the method body that will be executed when \keyword{m()} is invoked will the one found in class \className[B]. So, are we able to derive a predicate for \keyword{(new Sub()).upcast(new B(new A()))} that will allow the call to \keyword{m()} be typed?

In order to do this, we must find a predicate (assignable to the object \keyword{new Sub()}) for the \keyword{upcast()} method such that the result is the predicate mentioning \keyword{m} that we desire. This will be of the form \mkPredicate{\methodName[upcast]{:} `j \selfTypeOf \NPred \rightarrow \NPred} since the \keyword{upcast} method simply returns its argument. However, in addition to \NPred mentioning \methodName[m], it \emph{must} be the case that both $\NPred \ele \Lang(\className[A])$ \emph{and} \PAss{\PEnv}{\keyword{new B(new A())}}{\className[A]}{\NPred}. Here we come to heart of the matter: in order to derive a predicate describing the method \keyword{m} which we can assign to \keyword{new B(new A())}, we must look at the body of \keyword{m} \underline{in \className[B]}. This method body refers to the field \fieldId[f] in the receiver (\keyword{this}), and thus any predicate which we derive must also mention \fieldId[f]. However, since \fieldId[f] is not visible in the type \className[A] any such predicate will not be in the language of \className[A]. We find that there is no predicate \NPred which satisfies the necessary criteria and so we will not be able to assign a (non-trivial) predicate to expression (1) even though we can do so for its normal form, expression (4): e.g. \PAss{\emptyset}{\keyword{new A()}}{\className[A]}{\emptyPredicate} by using rule (\Rule{p-newO}).

Given that predicate languages are an essential element for the predictive abilities that we desire, the solution to the expansion problem will have to consist in modifying the definition of predicate languages to make them more permissive. In the example discussed above, we required a predicate to mention members which were not visible in the class of the language to which it belonged. Clearly, we must be able to allow predicates to contain such information, but only in the cases where it is necessary for expansion to hold since we still require that predicate languages make a statement about what is visible in a class and what is not.

\section{Conclusions and Future Work}
\label{section:Conclusions}

We have presented a predicate (type) system for \theLanguage, a variant of \FeatherweightJava, and shown that our predicates describe semantic properties of expressions. Our system thus has more expressive power than traditional type systems for Java. We see our results as important initial steps along the road to building not only semantic models for object oriented programming, but also practical analytic systems. A key development towards this aim will be to extend our system to a \emph{stateful} programming model, akin to Middleweight Java \cite{BiermanParkinsonPitts03}. Another objective of immediate concern to us is that of addressing the issues discussed in \textsection \ref{sec:Completeness} and achieving subject expansion.

%Unlike the intersection systems for the \LC, our system, while sound, is not completely expressive since predicable approximants may exist for an expression to which we cannot assign those same predicates. To correct this imbalance we need to modify our system to exhibit a \emph{subject expansion} property. While this may easily be achieved by discarding our notion of predicate language, that would destroy the semantic underpinning of our system (i.e.~the approximation result). This issue goes right to the heart of the object oriented paradigm since the failure of subject expansion in our system lies in the \emph{dynamic dispatch} feature of OO: an expression of type \Clss may reduce to an object of a \emph{subclass} of \Clss, which may override some method in \Clss and, in doing so, refer to new fields \emph{invisible} in \Clss. Any predicate describing this method's execution must refer to these new fields, yet remain in the language of \Clss for subject expansion to hold.

%\clearpage
\bibliographystyle{plain}
% \bibliography{bibliography1}

\end{document}